\documentclass[prb,preprint]{revtex4}
\usepackage{amsmath}
\usepackage{graphicx}

\begin{document}

\title{Comment on the cosmological constant and a gravitational alpha }

\author{Ronald J. Adler}
\email{adler@relgyro.stanford.edu}
\affiliation{Hansen Experimental Physics Laboratory, Gravity Probe B Mission, Stanford University, Stanford, California 94309}
\affiliation{Department of Physics and Astronomy, San Francisco State University, San Francisco, California 94132}


\begin{abstract}
We call attention to a simple analogy between atomic physics and cosmology. Both have two characteristic length scales. In atomic physics the lengths are the Compton wavelength of the electron and the Bohr radius; the ratio of these two lengths is the fine structure constant, $\alpha=7.30\times10^{-3}$. In cosmology we take the lengths to be the Planck length and the de Sitter radius divided by $\sqrt 3$; the ratio of these two lengths is about $\alpha_g=1.91\times10^{-61}$, which we suggest should be called the gravitational fine structure constant. There is also a basic energy ratio in atomic physics, the ratio of the hydrogen atom binding energy to the electron rest energy, which is equal to ${\alpha^2}/2$. The analogous energy ratio in cosmology is the ratio of the dark energy density (described in terms of the cosmological constant) to the Planck energy density, which is equal to $(1/8\pi)\alpha_g^2$. The long-standing problem of the nature of the dark energy and its small density is obviously equivalent to understanding the extraordinarily small value of  $\alpha_g$. We further emphasize that our observational knowledge of dark energy, which is consistent with the cosmological constant interpretation, is entirely on the cosmological scale, so we know essentially nothing about the nature of dark energy on a smaller and presumably more fundamental scale. 
\end{abstract}

\maketitle

\section{Introduction}
The modest purpose of this paper is to call attention to an analogy between an old problem and a new problem in physics. The old problem was to understand the nature and energy levels of atoms, the hydrogen atom in particular, which was a center of attention in the early twentieth century. \cite{1} The new problem is to understand the nature of the dark energy, which is a center of attention now in the early twentieth-first century. \cite{2}

Both problems involve two characteristic lengths, whose ratios are thus dimensionless constants. The relevant constant in atomic physics is the fine structure constant  $\alpha=7.30\times10^{-3}$, which plays a fundamental role in atomic physics and quantum electrodynamics, QED. \cite{3} The analogous constant in cosmology is $1.91\times10^{-61}$. It plays a fundamental role in cosmology and we suggest it should be viewed as a gravitational fine structure constant $\alpha_g $. \cite{4}

In section 2 we recall the length scales of atomic physics and cosmology and display the analogy and ratios noted above.  In section 3 we do the same for the energy scales, assuming that dark energy may be interpreted, either exactly or approximately, in terms of a cosmological constant. In section 4 we summarize the analogy between atomic physics and cosmology. We further note  that the current observational evidence about dark energy is limited to the cosmological scale and we remain quite ignorant of its properties on any smaller scale; in particular we have no observational evidence that dark energy is uniform or time independent on a smaller scale, as the cosmological constant interpretation of dark energy would imply.\cite {5,6}  Indeed we do not have observational evidence that dark energy interacts only gravitationally with the other contents of the Universe. \cite {5} 

\section{Length scales}
Atomic physics contains two characteristic lengths, the electron Compton wavelength, $\lambda_e$, and the Bohr radius, $a_B$. Both lengths depend only on intrinsic properties of the electron, and are given in terms of the electron mass $m_e$, charge $e$, and Planck's constant  $\hbar$ by
\begin{equation}
\label{1}
\lambda_e=\hbar/m_e c=3.86\times 10^{-13}m,  \ a_B=\hbar^2/m_e e^2=5.29\times 10^{-11}m.
\end{equation}	
The ratio of these lengths is the fine structure constant 
\begin{equation}
\label{2}
\frac{\lambda_e}{a_B}=\frac{(\hbar /m_e c)}{(\hbar^{2}/m_e e^2)}=\frac{e^2}{\hbar c}=\alpha=7.30\times 10^{-3}.
\end{equation}		
The constant $\alpha$ plays a fundamental role in atomic physics and quantum electrodynamics, QED, and is probably the most famous dimensionless constant in physics. For example calculations in atomic physics routinely give energy levels in terms of powers of $\alpha$, and QED amplitudes are generally expressed in terms of powers of $\alpha$. The deeper meaning of $\alpha$ involves its interpretation and value as a running coupling parameter in the unified electroweak theory and possibly in a grand unified theory, and remains of basic research interest. \cite{7}

Cosmology also contains two characteristic lengths; the first is the Planck length, which is related to Newton's gravitational constant by 
\begin{equation}
\label{3}
L_P=\sqrt{G\hbar/c^3}=1.62\times 10^{-35}m.
\end{equation}
Note that since the Planck length contains $\hbar$ it cannot be associated with purely classical gravity. \cite{8}

To introduce the second length we make the standard assumption that dark energy on the cosmological scale is well-described by the cosmological constant; that is we use the $\Lambda$CDM model for the present Universe. \cite{2}  The cosmological constant $\Lambda$ has the dimension of an inverse length squared. \cite{9} That length, $1/\sqrt{\Lambda}$, is related to the de Sitter radius by 
\begin{equation}
\label{4}
\frac{R_d}{\sqrt 3}=\sqrt{\frac{1}{\Lambda}}.
\end{equation}									
The numerical value of the de Sitter radius is conveniently obtained from its relation to the measured Hubble parameter $H=73(km/s)/Mpc$ and corresponding Hubble distance $L_H=1.27\times 10^{26}m $ and the measured dark energy density ratio $\Omega_\Lambda=0.75$  via the relation
\begin{equation}
\label{5}
R_d=L_H/\sqrt{\Omega_\Lambda}=1.46\times10^{26} m.
\end{equation}	 
Since dark energy should ultimately dominate the universe, that is $\Omega_\Lambda\rightarrow1$, the $\Lambda$CDM model approaches the de Sitter model and the Hubble distance approaches the de Sitter radius. \cite{9} 

The ratio of the Planck length in (3) to the de Sitter radius divided by $\sqrt 3$ in (5) is a rather elegant expression containing the constants $\hbar$, $c$, $G$ and $\Lambda$, which has an extraordinarily small value, 
\begin{equation}
\label{6}
\frac{L_P}{(R_d/\sqrt 3)}=\frac{\sqrt{G\hbar /c^3}}{\sqrt{1/\Lambda}}=\sqrt{G\hbar\Lambda / c^3}=\alpha_g=1.91\times10^{-61}
\end{equation}
Because of the analogy between the ratios (2) and (6) we suggest that it is reasonable to call the ratio in (6) the gravitational fine structure constant or $\alpha_g$.  Why the Universe should contain two such different characteristic lengths with such a small ratio remains a mystery. \cite{2}

\section{Energy scales}

Atomic physics has two characteristic energies, the rest energy of the electron $E_e$, and the binding energy of the hydrogen atom $E_H$, 
\begin{equation}
\label{7}
E_e=m_e c^2=0.511MeV,  \ E_H=m_e e^4/2\hbar^2=13.6eV.
\end{equation}
Their ratio is equal to half the square of the fine structure constant, 
\begin{equation}
\label{8}
\frac{E_H}{E_e}=\frac{m_e e^4/2\hbar^2}{m_e c^2}=\frac{1}{2} \Big( \frac{e^2}{\hbar c} \Big)^2
=\alpha^2 /2
=2.66\times10^{-5}
\end{equation}

Cosmology also has two characteristic energy scales, the Planck energy density $\rho_P$, and the density of the dark energy $\rho_\Lambda$. The Planck energy density is defined as the Planck energy, 
\begin{equation}
\label{9}
E_P=\frac{\hbar c}{L_P}=\sqrt{\hbar c^5/G}=1.22\times 10^{19} GeV, 
\end{equation}
divided by the Planck volume ${L_P}^3$, 
\begin{equation}
\label{10}
\rho_P=\frac{E_P}{{L_P}^3}=\frac{c^7}{\hbar G^2}=2.89 \times 10^{123} \ GeV/m^3.
\end{equation}	
To obtain an expression for the dark energy density in terms of the cosmological constant we recall that the cosmological term in the general relativity field equations may be interpreted as a fluid energy momentum tensor of the dark energy according to
\begin{equation}
\label{11}
\Lambda {g^\mu}_\nu=(8\pi G/c^4){T^\mu}_\nu (de),
\end{equation}
so the dark energy density $\rho_\Lambda={T^0}_0 (de)$  is given by
\begin{equation}
\label{12}
\rho_\Lambda=\Lambda c^4/8\pi G.
\end{equation}
To evaluate numerically the density $\rho_\Lambda$ we express it in terms of the Planck length and energy in (3) and (9) and the de Sitter radius in (4) and obtain
\begin{equation}
\label{13}
\rho_\Lambda=(3/8\pi) \Big(\frac{E_P}{L_P {R_d}^2}\Big)=4.21 \ GeV/m^3.
\end{equation}
The ratio of the energy densities in (13) and (10) is thus the extremely small quantity 
\begin{equation}
\label{14}
\frac{\rho_\Lambda}{\rho_P}=(3/8\pi)\Big(\frac{L_P}{R_d}\Big)^2=(1/8\pi){\alpha_g}^2=1.45\times 10^{-123}
\end{equation}
Eq. (14) is the cosmological analog of the atomic energy ratio in (8). 

Some authors consider the small value of the ratio (14) to be arguably one of the most mysterious problems in present day physics. The understanding of atomic structure required the discovery of the fundamental dynamical constant $\hbar$. Viewed in this way the cosmological analog of $\hbar$ is $\Lambda$, but any dynamical role it may play is not yet apparent. 

It is amusing to note that in the presence of two length scales, and their dimensionless ratio, dimensional analysis becomes problematic; a dimensional estimate can contain an arbitrary function of the ratio, for example a power or a logarithm. In the case of cosmology it is clear that dimensional estimates, with two disparate length scales, may be much worse than useless.   

\section{summary and further comments}

We have pointed out a simple analogy between atomic physics and cosmology that suggests $\alpha_g=\sqrt{G\hbar \Lambda /c^3}=1.91\times10^{-61}$ may be thought of as a gravitational fine structure constant. In view of the dynamical role played by the fine structure constant $\alpha$ in atomic physics the analogy may suggest, albeit rather vaguely, that dark energy (i.e. the cosmic vacuum) has a more interesting structure than if described entirely by the cosmological constant; emergent gravity theories and quintessence theories offer examples of such structure. \cite{6,10} We should also stress our present state of observational ignorance of dark energy at a scale less than cosmological. In particular we have no observational evidence that dark energy is uniform or time independent on a small scale. The possibility even exists that dark energy could have non-gravitational interactions, as dark matter is presently presumed to have via the weak force. \cite{5}

\section{acknowledgments}
We thank James Bjorken for stimulating and provocative discussions on the nature of spacetime and gravity, and Martin Perl and Pisin Chen for many interesting discussions on dark energy and cosmology. 
We also thank Mustafa Amin for providing interesting conjectures on the nature and role of quintessence.

\end{document}